\documentclass[preprint,final,5p,times,twocolumn]{elsarticle}
\usepackage{rotating,color,subfigure,amssymb}
\usepackage{alphalph}
\usepackage{amsmath}
\usepackage[T1]{fontenc}
\usepackage{amssymb}
\journal{Physics Letters B}
\begin{document}
                                       
\begin{frontmatter}

\title{ Novel Six-Quark Hidden-Color Dibaryon States in QCD}
\date{\today}

\author[PITue]{M.~Bashkanov}\corref{coau}\ead{bashkano@pit.physik.uni-tuebingen.de}   
\author[SLAC]{Stanley J.~Brodsky} 
\author[PITue]{H.~Clement}    

\address[PITue]{Physikalisches Institut, Eberhard--Karls--Universit\"at 
 T\"ubingen, Auf der Morgenstelle~14, 72076 T\"ubingen, Germany}
\address[SLAC]{SLAC National Accelerator Laboratory Stanford University}

\cortext[coau]{Corresponding author }

\begin{abstract}
The recent observation of a hadronic resonance $d^*$ in the proton-neutron
system with isospin $I = 0$  and  spin-parity $J^P = 3^+$ 
 raises the possibility of producing other novel six-quark dibaryon
 configurations allowed by 
QCD.  A dramatic example of an exotic six-quark color-singlet system is the
charge $Q=+4$, isospin $I=3$, $I^z=+3$   $|uuuuuu>$ state which couples
strongly to $\Delta^{++}$ + $\Delta^{++} .$ 
The width and decay properties of such six-quark
resonances could be regarded as manifestations of  "hidden-color" six-quark
configurations, a first-principle prediction of QCD -- SU(3)-color gauge
theory for the deuteron distribution amplitude. Other implications and possible future experiments are discussed. 
\end{abstract}



\end{frontmatter}


\section{Introduction}
Because of color confinement, one expects that virtually any color-singlet
hadronic configuration of quarks and gluons can form either bound states or
resonances. In addition to the familiar $q \bar q$ mesons, $qqq$ baryons, the
$g g$ and $ggg$ glueball states~\cite{ochs}, as well as nuclei, color
confinement can lead to  $q \bar q q \bar q$  "tetraquark"
systems~\cite{Karliner:2013dqa} such as the charged $Z_c(c \bar c u \bar
d)$\cite{bes,belle1} and possibly $qqqq \bar q$ "pentaquark"
states~\cite{Fock}.    Mesonic
nuclei~\cite{mosk,yokota,Oset1,Gal1,Nagahiro,Bayar,Kaskulov}   and
nuclear-bound quarkonium ~\cite{Brodsky:1989jd,Luke:1992tm, bel} are also
possible. 
Resonances in the $\bar q \bar q \bar q q q q$ channel
just below the $B \bar B$ threshold could explain the anomalously large
rates~\cite{Baldini:2007qg} seen in $e^+ e^- \to p \bar p, n \bar n, \bar
\Lambda \Lambda$ at threshold. The anomalously large transverse spin-spin correlation $A_{NN}$ observed in 
large-angle proton-proton elastic scattering near the strangeness and charm thresholds~\cite{Court:1986dh} could be explained by the effects of  $|uud uud Q \bar Q>$ 
baryon number $B=2$ resonances in the $J=L=1$ $pp$ $s$-channel~\cite{Brodsky:1988mol,Brodsky:2012mol}.

Understanding the mechanisms underlying confinement in QCD is among the most
fundamental questions in hadron physics.  
In the case of heavy quarks, the potential evidently can be identified with
gluon exchange, in analogy with the Coulomb forces which bind atoms.    
The  potential underlying light-quark interactions is however much more
complex --  such as flux-tube exchange~\cite{Isgur:1984bm}  and other string-like
forces~\cite{Makeenko:2011gj} built from multi-gluon exchange.  It has
recently been shown that the effective confining $q \bar
q$ potential in the frame-independent QCD light-front (LF) Hamiltonian has a
unique form~\cite{Brodsky:2013ar} if one maintains conformal symmetry of the QCD action,  
The resulting meson eigensolutions of the resulting light-front Schrodinger
equation include a zero-mass pion in the chiral $m_q\to 0$ limit, and linear
Regge trajectories ${M}^2(n,L) \propto n+L$  with the same slope  in the 
radial quantum number $n$ and orbital angular momentum $L$. 
In the case of light baryons, the confining potential could mimic the $q \bar
q$ form as a  two-body quark-diquark interaction in a light-front Dirac
equation~\cite{deTeramond:2011qp,Brodsky:2013sc}, or take the form of a
three-body force such as a $Y$ junction~\cite{Takahashi:2000te} between the
valence quarks.   The LF Dirac equation based on quark-diquark interactions
with the same potential as $q \bar q$ accounts well  for the  measured light
baryon spectrum~\cite{Brodsky:2013sc} 

The possible mechanisms underlying confinement multiply as the number of
quarks and gluon constituents in a hadronic system  increase.  A key question is whether such states
bound by fundamental QCD interactions or do the constituents always cluster as
color-singlet subsystems?  In the case of nuclei,  the quark constituents
evidently cluster as color-singlet nucleons bound by virtual meson exchange,
the analog of covalent binding in molecular physics due to quark interchange
or exchange.   
When there are no covalence quarks in common, QCD also predicts attractive
multigluonic van der Waals forces which are dual to glueball
exchange.   The attractive QCD van der Waals potential leads to the prediction
of bound states of heavy quarkonium to heavy
nuclei~\cite{Brodsky:1989jd,Luke:1992tm,YI:2013vba}. 
However, there are also rare configurations in which other multiquark color
configurations ("hidden color"~\cite{br}) can enter.

There are several possible interpretations~\cite{Mahajan:2013qja} for the
dominant internal structure of the positively charged $Z^+_c(4025), $ which can be 
identified as a $|c \bar c u d>$ color-singlet tetraquark bound state. 
The $Z_c$ could be considered an example of a bound state of $c \bar c$
quarkonium with a light $u \bar d$ meson bound by gluon exchange,
corresponding to "disconnected contributions" in lattice gauge theory
simulations ~\cite{Guo:2013nja}); or a $ D^* \bar D^*$ hadronic
molecule~\cite{Voloshin:1976ap,Karliner:2013dqa,Qiao:2013dda} such as $ 
D^*(c \bar u) \bar D^*(\bar c d)$ clusters bound by meson exchange. Other
color-confining interactions between higher-color multiquark representations  
may also dominate~\cite{Weinberg:2013cfa,Lebed:2013aka}.

The possibility of exotic six-quark $qqqqqq$ dibaryonic "hexaquark" states was
first  proposed by  F. J. Dyson and N. Xuong~\cite{dys} in 1964, just a half a
year after Gell-Mann's publication of the quark
model~\cite{Gell-Mann}.  However, this topic received intensive attention only
after  Jaffe's proposal\cite{Jaffe} of the so-called "H dibaryon", a $|uuddss>$ state
corresponding asymptotically to a bound $\Lambda\Lambda$ system. This
hypothesis initiated a worldwide activity of theoretical predictions for
dibaryon states with and without strangeness -- as well as numerous experimental
searches.  Despite numerous claims,  no 
established dibaryon candidate has  emerged.  For a recent report concerning
the experimental H dibaryon search see {\it e.g.} Ref. ~\cite{belle}.
However, there has been renewed interest in such states, in part because
lattice QCD calculations are now becoming available~
\cite{beane,inoue,beane1a,beane1b,beane1c,hal,aoki}.

As we shall show in this paper, the discovery of six-quark states would
provide a novel extension of the domain of hadronic states in QCD, and  the 
experimental verification of such dibaryon states may well be
possible in the near future. 

The most familiar six-quark state is the isospin-zero 
$|uud ddu>$ deuteron;  in fact, the wavefunction of the deuteron has
novel properties in QCD.   Five distinct color-singlet configurations of six
color triplets $3_C$ can form a color singlet in $SU(3)$ color, only one of
which corresponds to the usual $p n$ configuration.  When one probes the
light-front  
wavefunctions of the deuteron where all of the six quarks have small relative
separation, as in the deuteron form factor at high moment transfer or in
photodisintegration $\gamma d \to n p$ at high transverse momentum,  the five
"hidden-color configurations" of the deuteron mix due to gluon exchange and
become equal in magnitude at asymptotic $Q^2\to \infty $ ~\cite{br}.  
For example, the observed $Q^{10} F_d(Q^2)$ scaling~\cite{Brodsky:1976mn} of
the deuteron $\sqrt A(Q^2)$ form factor at high $Q^2$~\cite{Rock:1991jy} is
dominated by  hidden-color configurations. This result can be derived 
by applying ERBL evolution~\cite{bl,er} to the five-component deuteron
distribution amplitude $\phi_d(x_i,Q)$. 
The color-singlet states of the deuteron wavefunction also couple to a virtual
$\Delta^+ \Delta^0$ state~\cite{Frick:2001ms}.

The most dramatic example of an exotic six-quark color-singlet system is the
charge $Q=+4$, isospin $I=3$, $I^z=+3$   $|uuuuuu>$ state, as
originally proposed by Dyson and  Xuong~\cite{dys}.
The Fermi-Dirac statistics of the  color-triplet  $u$quark only allows one
color-singlet six-quark  configuration with zero orbital angular momentum:  
$|u^\uparrow_R u^\uparrow_Y u^\uparrow_B u^\downarrow_R u^\downarrow_Y
u^\downarrow_B>$.  
The set of seven $I^z=3,2,1,0,-1,-2,-3$ states ranging from $Q=+4$ to $Q=-2$
are then obtained by applying the isospin-lowering operator.  As a first approximation, one can estimate
their masses $\simeq 2.4$ GeV  by considering these states as effective
$\Delta\Delta$ bound states;  e.g. the  $|uuuuuu>$ state can be considered as a
bound-state of two $ I^z= 3/2$  $\Delta^{++}$ isobars 
     -- for model predictions of its mass see, {\it e.g.}
     Refs. \cite{dys,kamae,gold,muld,aer,malt,mota}. 
We do not expect 
major Coulombic corrections to its dibaryon properties from the high $Q=+4$ charge 
since one does not observe significant charge-related effects in $nn-pp$ or in
$\Delta^{++}$-$\Delta^0$ systems. 

In this paper we will review the present evidence for dibaryon states and discuss future strategies for
detecting such six-quark states.  A typical example is the study of $p p
\to  \pi^- \pi^- X$, where the recoil system $X$ could display a charge $Q=+4$
resonance peak in the $X$ missing mass.   One can also look for $\Delta^{++}
\Delta^{++}$ resonance decay as an enhancement in the rate of the exclusive
channel measurement $p p \to  \pi^- \pi^- \Delta^{++} \Delta^{++}$.   The
enhancement could appear below the nominal two-isobar mass, indicating a
possible $\Delta \Delta$ bound state phenomenon. 

\section{Recent Experimental Evidence for a $\Delta\Delta$ Resonance}

A pronounced resonance structure has recently been observed in $p n$
collisions  leading to two-pion production in the reactions $pn \to
d\pi^0\pi^0$~\cite{mb,MB}, $pn \to d \pi^+\pi^-$~\cite{MBC}, $pn \to pp
\pi^-\pi^0$~\cite{ts} and possibly also in  $pn$ elastic scattering, in
particular in the total cross section and in the analyzing power
\cite{MBEl}. For the not yet measured reactions $pn \to pn \pi^0\pi^0$ and $pn
\to pn \pi^+\pi^-$ exist predictions for the size of the expected resonance
effect \cite{colin,oset}.

The measured parameters for this resonance structure, called henceforth
$d^*$, are  $I(J^P) = 0(3^+)$ with mass $M$ = 2.37~ GeV and width
$\Gamma$~=~70~MeV~\cite{MB,MBC,ts}.  Dalitz plots indicate 
that $d^*$ dominantly decays via an intermediate $\Delta-\Delta$
system. However, the mass of this resonance is about 90 MeV below the nominal
mass 2$m_{\Delta}$ of a
$\Delta\Delta$ system, and its width is about three times smaller than
that of a $\Delta\Delta$ system formed by conventional $t$-channel meson
exchange or quark interchange arising within the $NN$ collision processes.
The interchange of quarks of the same flavor~\cite{Gunion:1973ex} has been
shown to dominate  
hadron-hadron elastic scattering amplitudes in the hard-scattering fixed
$\theta_{CM}$ scattering domain~\cite{White:1994tj}.

We conclude from such observations that $d^*$ must be of an unconventional
origin, possibly indicating a genuine six-quark nature. With the predominant
decay of $d^*$ being $d^* \to \Delta\Delta$ 
($BR(d^* \to \Delta\Delta)/BR(d^* \to pn$) = 9:1), one could naively expect
$d^*$ to be a so-called  a "deltaron" denoting a deuteron-like
bound state of two $\Delta$s.   
However, the narrow width of $d^*$ contradicts
this simple assumption. A deltaron would need to have 90 MeV binding
energy, {\it i.e.} 45 MeV per $\Delta$, which would lead to a reduction of
width from $\Gamma_{\Delta\Delta}$ = 230 MeV to 
$\Gamma_{\Delta\Delta}$ = 160 MeV, using the known momentum dependence of the
width of the $\Delta$ resonance. This is more than twice what is observed.

On the other hand, if $d^*$ is a genuine six-quark dibaryon state, we need
to understand its large coupling $d^* \to 
\Delta\Delta$.   This can be explained if one assumes the  $d^*$
is dominated by a "hidden-color" six-quark state. Hidden-color six-quark
states  are a rigorous first-principle prediction of  SU(3) color 
gauge theory. Six quark color-triplets $3_C$ combine to five different
color-singlets in QCD,  -- and as shown in Ref.~\cite{br}, will significantly
decay to $\Delta \Delta$. 

According to M. Harvey~\cite{har} there are only two
possible quark structures for an $I(J^P) = 0(3^+)$ resonance in the two-baryon
system: 

 $\vert \Psi_{d^*} \rangle = \sqrt{\frac{1}{5}} \vert
\Delta\Delta \rangle + \sqrt{\frac{4}{5}} \vert 6Q \rangle$ and 

$\vert \Psi_{d^*} \rangle = \sqrt{\frac{4}{5}} \vert \Delta\Delta \rangle -
   \sqrt{\frac{1}{5}} \vert 6Q \rangle$.

Here $\Delta\Delta$ means the asymptotic $\Delta\Delta$ configuration and $6Q$
is the genuine 
"hidden color" six-quark configuration. The first solution denotes a 
{$S^6$} quark structure (all six quarks in the S-shell), the second one a 
{$S^4P^2$} configuration (4 quarks in the S-shell and 2 quarks in the
P-shell). The quark structure with the large $\Delta\Delta$ coupling
would correspond to a deltaron and can be excluded. Thus it is natural to  
assign the observed $d^*$ resonance to the $S^6$ six-quark predominantly
"hidden color" state, thus  providing an explanation for its narrow
decay width.

   The above ansatz for the $d^*$ wavefunction and the decay of $d^*$ via the
   $^7S_3$ configuration of the intermediate $\Delta^+\Delta^0$ system
   imply that the resulting nucleon and pion pairs are either both in I = 0 or
   both in I =1 isospin states. The first possibility is realized in the
   experimentally studied channels $d\pi^0\pi^0$ \cite{MB}and $d\pi^+\pi^-$
   \cite{MBC}, and the second possibility can be studied in the $pp\pi^0\pi^-$
   channel, where a signal for the $d^*$ has been observed, too \cite{ts}. In
   the I = 1 case the pion pair must be in relative $p$-wave in order to
   comply with Bose statistics; in addition,  the nucleon pair must be in a
   relative $p$-wave or higher in order to obtain the required spin and parity
   of the $d^*$. Higher orbital angular momenta cannot be excluded at present,
   requiring a more complicated wavefunction than given above.

Due to its quantum numbers, the $d^*$ state must be fully symmetric in spin,
color, and angular momentum as well as fully antisymmetric in isospin. Due to
this particular feature, Ref.~\cite{gold} claims that any model based on
confinement and effective one-gluon exchange leads to the prediction of the
existence  of a non-strange dibaryon with $I(J^P) = 0(3^+)$, the 
"inevitable non-strange dibaryon".  In fact, many
groups~\cite{dys,kamae,gold,muld,aer,malt} predicted such a state at
similar  mass. 
     Very recently Gal and Garcilazo succeeded to predict such a deeply bound
     $\Delta\Delta$ state at the experimentally observed mass in a fully
     relativistic three-body calculation based on hadron dynamics
     \cite{Gal:2013dca}.

It is remarkable that the first such calculation published by
Dyson and Xuong ~\cite{dys} appears now to be
quite precise in the prediction of the $d^*$ mass. In the nomenclature of
Ref.~\cite{dys} , the $d^*$ has the notation $D_{03}$, where the indices
$(03)$ denote the  
isospin $ I=0 $ and spin $J=3$ of the dibaryon. To predict the mass of
$D_{03}$  Dyson and  Xuong identified the $D_{01}$ state with the $^3S_1$
deuteron ground-state and the $D_{10}$ with the $^1S_0$ virtual state
(unbound by 66 keV only~\cite{Wir}),  
which is known to contribute to the nucleon-nucleon
final-state interaction. 
These two states are also currently being used to check the reliability of lattice
calculations for the H-dibaryon~\cite{beane1a,beane1c,aoki,ukawa,ukawa1}.

Most  quark models predict~\cite{dys,kamae,gold,muld,aer,malt} that in addition to $d^*$
one should have also a state with mirrored quantum numbers for spin and
isospin, {\it i.e.} $I(J^P) = 3(0^+)$ at a similar mass. Such a state, which in
the notation of Ref.~\cite{dys} is $D_{30}$, would be
symmetric in isospin, color, angular momentum and antisymmetric in
spin. Due to its isospin $I = 3$, it cannot decay into $NN$ or $NN\pi$, but
only into the $NN\pi\pi$ channel. Thus if such a state has a mass close to
that of $d^*$,  its width must be even smaller than that of $d^*$.

According to Ref. \cite{dys}, both $d^*$ and $D_{30}$ belong to multiplets
of dibaryons, the first one is assigned to  an antidecuplet and the second one
to a 28-plet. Thus, given  the existence of  the $d^*$,  one should expect a 
number of strange dibaryons. The three corners of the possible 
28-plet look truly exotic: 6u quarks, 6d quarks, 6s quarks. In each
of these cases the quarks occupy all possible states. The 6s quark state can
be considered as a strange droplet and could play an important role in
astrophysics regarding the nuclear equation of matter in the core of neutron
stars. Recent calculations on $\Omega\Omega$ (6s quark
state)~\cite{zhang,wang,buch} display  a range of results --- from $100~MeV$
binding to an unbound state.

\section {Experimental Strategies}

The existence of novel dibaryon states still awaits definitive experimental
confirmation or  exclusion. Thus we will discuss in the following a number of
possible experiments and strategies for producing the  
charge-1 $d^*$ and charge-4 $D_{30}$, such as electro- and photo-production of
$d^*$ on a deuteron:
 $\gamma d \to d^* \to d \pi^0\pi^0$. A suitable place to perform such an
experiment appears to be MAMI at Mainz due to its high beam intensity
and good neutral particles detection capabilities of the Crystal
Ball experiment. Such a reaction should preferably go via
photon coupling to the deuteron's six-quark component and will allow to fix
the transition from the six-quark component in the deuteron to the one of
$d^*$. The reaction $\gamma d \to d^* \to d \pi^+\pi^-$ is less favorable due
much higher background rates ~\cite{fix}. 
     For a first calculation of the $d^*$ electroproduction see
     Ref. \cite{wong}.

    Due to the expected small cross sections, such experiments require
    sufficient beam intensity as well as large detector acceptance
    for photons and efficient particle identification in order to discriminate
    deuterons from protons. Several detector setups fulfill these conditions:
    the Crystal Ball at MAMI/Mainz ~\cite{CBall}, the Crystal Barrel
    ~\cite{CBarr} and BGO-OD at Elsa/Bonn. The use of a polarized photon beam
    and a polarized deuteron target at the Crystal Ball and Crystal Barrel
    experiments can provide further suppression of the conventional
    background. In addition, the Crystal Ball, in combination with a nucleon
    recoil polarimeter ~\cite{Watts}, can investigate the $\gamma
    d \to d^* \to \overrightarrow{p} \overrightarrow{n}$ reaction where a
    sign of the $d^*$ resonance was observed in the seventies
    ~\cite{kamae,Ikeda}.

With the knowledge of the $dd^*\gamma$
coupling one can estimate possible cross sections for the
production of other antidecuplet members in reactions like $\gamma d
\to d^*_s + K^+\to \Delta \Sigma^* + K^+$. Such reactions could be measured 
at JLab.  Another possibility to produce the strange partners of $d^*$ would
be the study of kaon-induced reactions  of the kind $K^-d \to d^*_s \to \Delta
\Sigma^*$ as could be conducted at JPARC.

Accessing the members of the 28-plet appears to be much more complex. Most
prominent here is $D_{30}$ with charge $Q = +4$ (six u-quarks). The dedicated
decay channel of such a state is $D_{30} \to pp \pi^+ \pi^+$ which can be
triggered with  high selectivity.  However, the production of such a state is
challenging. One may be able to produce it in $pp$ collisions;  however, in
order to reach the  $I=3$ state, one needs to produce in addition two associated
negative pions $pp \to D_{30} \pi^- \pi^- \to (pp \pi^+ \pi^+)  \pi^-
\pi^-$. To perform such a reaction in the energy region of interest, one needs
a rather high beam energy of $T_p$ =1.7-2~GeV which is available at COSY and
JPARC.  However, the $pp\pi^+\pi^+\pi^-\pi^-$ channel will be
highly contaminated by conventional $N^*$ and $\Delta$ excitations. 

Another important way to identify  the $D_{30}$ is its production in nuclei,
{\it e.g.} on carbon by the reaction
$\gamma C^{12} \to pp \pi^+ \pi^+ X$ below the 4$\pi$ threshold at JLab,  or
similarly using proton or pion beams in reactions  such as $p ^{12}$C$ \to
pp \pi^+\pi^+ X$ and $\pi^+$$ ^{12}$C$ \to pp \pi^+ \pi^+ X$. In all such
reactions the conventional background due to associated meson production
production needs to be effectively suppressed. 
    Magnetic separation of positively and negatively charged pions is a
    prerequisite for measuring such reactions due to the much higher $\pi^+
    \pi^-$ rate. The CLAS detector ~\cite{Clas} at JLab with toroidal magnetic
    field, large acceptance and high momentum resolution would be a very
    suitable place to perform such experiments. 

The detection of the $Q = +4$  $D_{30}$ resonance would help to constrain the
properties of the "strange droplets", the $\Omega \Omega$ states, and thus
simplify its search in heavy-ion collisions.
 
Another place to look for both  the $D_{03}$ ($d^*$) and $D_{30}$ resonances is
to search in quarkonium decays. The high mass of dibaryonic resonances
excludes charmonium decays;  however, bottomium decays measured at
B-factories appear to be promising.  The observation  of the  $d^*$  looks
particularly straightforward: due to its isospin $I=0$ one does not necessarily need
to search for $\Upsilon \to \bar{d^*}d^*$;  the search for $\Upsilon \to \bar{d}d^*$
would be sufficient. The branching ratio of $BR(\Upsilon \to \bar{d}+X) = 2.86
\times 10^{-5}$~\cite{BR} appears to be large enough to search for the
reaction $\Upsilon \to \bar{d} d^*$ or  $\bar{d^*} d \to d\bar{d}(\pi\pi)_{I=0}$. 
This simple possibility is forbidden for the $D_{30}$ because of its
isospin.  One could produce  the $D_{30}$ paired with  $\overline{D}_{30}$
having in minimal configuration  $\Upsilon \to \overline{D}_{30} D_{30} \to
(\bar{p}\bar{p}\pi^-\pi^-) (pp\pi^+\pi^+)$. Unfortunately, this channel will
contain large contamination from the production of conventional $N^*$ and
$\Delta$ resonances and their antimatter analogs. However,   one
can extract not only the mass and width of resonances in this way, but also its time-like form-factor.  The extraction of the space-like form-factor for
such a state appears to be impossible at the present level of experimental
capabilities,  so distinguishing between molecular-type and genuine dibaryon will
be challenging.  

To our knowledge dibaryon channels have not yet
been looked for at $e^+ e^-$ colliders;  however, the statistics of data already collected at BaBar
and Belle should be large enough to search for such resonances.
Recent publications on the search for H-dibaryon production in the
$\Upsilon$ decays by Belle~\cite{belle} may be a good starting point for the
search of other dibaryon candidates, including the ones discussed here.

Another important experimental option is the photoproduction or electroproduction of a dibaryon state on
a nucleon target in combination with associated anti-nucleon production, such as 
$\gamma p\to \bar{p} \pi^- \pi^-  D_{30}\to(\bar{p}\pi^-\pi^-) (p p\pi^+\pi^+)$,  a reaction  which could be investigated at the upcoming
12~GeV electron facility at  JLab. The advantage of such reactions is the
particularly simple triggering conditions  -- the essential signal for the dibaryon is provided by the antiproton trigger.  
Of course, as in the other cases discussed above,
one may encounter a high level of conventional backgrounds.

The triggering on antiparticles promises to be even better suited for strange
dibaryons. In case of strangeness $S = -1$, the tagging on the $\overline{\Sigma^*}$ allows one
to separate antidecuplet from 27-plet states.  Only a 27-plet $J=0$ state can be produced in combination with 
$\overline{\Sigma^{*-}}$ in the reaction $\gamma p\to \overline{\Sigma^{*-}}
+ (\Sigma^{*+} \Delta^{+})_{27}$, whereas with $\overline{\Sigma^{*+}}$ both
the 27-plet $J=0$ and antidecuplet $J=3$ states are possible in the process
$\gamma p\to \overline{\Sigma^{*+}} + (\Sigma^{*-} \Delta^{+})_{27,\bar{10}}$.
Similarly, one can search for double and triple-strange dibaryons in the
reactions $\gamma p\to
\overline{\Xi^{*+}} + (\Xi^{*-} \Delta^{+})_{27,\bar{10}}$ and $\gamma p\to
\overline{\Omega} + (\Omega \Delta^{+})_{27,\bar{10}}$.  in addition, tagging on
antiparticles can effectively suppress conventional backgrounds.

\section{Summary} 
The recent observation of a narrow hadronic proton-neutron resonance  $d^* $
with $I(J^P) = 0(3^+)$ and  mass $M = 2.37~ GeV $ raises the possibility of
producing other novel color-singlet six-quark dibaryon configurations allowed
by QCD.  A dramatic example would be the discovery of an exotic six-quark
$|uuuuuu>$ color-singlet system with charge $Q=+4$, isospin $I=3$, and
$I^z=+3$, a   state which couples strongly to $\Delta^{++}$ + $\Delta^{++} .$
The width and decay properties of such six-quark resonances could be regarded
as  a manifestation of either  a "hidden-color" six-quark configuration,
versus a  more conventional  interpretation as a $\Delta-\Delta$  (deltaron)
resonance. We have discussed a number of possible experiments where such a
state could be observed. 

\section{Acknowledgments}
We  are grateful to David Bugg, William Detmold, Avraham Gal, Johann
Haidenbauer, Christoph Hanhart, Marek Karliner, Eulogio Oset and Colin Wilkin
for helpful discussions.  This work  was supported by 
the Department of Energy contract DE--AC02--76SF00515, the BMBF (06TU9193) and
the Forschungszentrzum J\"ulich (COSY-FFE).    SLAC-PUB-15720.

\end{document}